\documentclass[allclo,epsf]{FBSart}
\usepackage{amsfonts}
\usepackage{amssymb}
\newcommand{\bi}{\bibitem}

\title{Reaction Mechanisms of the Proton - Deuteron Breakup Process at GeV Energies}

\author{N.B.Ladygina
$^{1}$
\thanks{\textit{E-mail address:}
ladygina@sunhe.jinr.ru},
A.V.Shebeko$^{2}$
\thanks{\textit{E-mail address:}
shebeko@fbsyst.kharkov.ua}}
\institute{$^1$Laboratory of High Energies,
Joint Institute for Nuclear Research,\\ 141980 Dubna, Russia,\\
$^2$NSC Kharkov Institute of Physics \& Technology, 61108 Kharkov, Ukraine.}

\runningauthor{N.\,B.\,Ladygina A.\,V.\,Shebeko}
\runningtitle{Reaction mechanisms of proton-deuteron breakup at 1 GeV}
\begin{document}

\maketitle
\begin{abstract}
The deuteron fragmentation by fast protons
has been studied both near the kinematics of
  quasi-free proton - proton scattering and far away
from it. We have concentrated on the interplay between different reaction
mechanisms associated with the antisymmetrization of the initial and final
states and rescattering contributions. A multiple-scattering-expansion
technique has been applied to evaluate  the reaction amplitude. An
essential element of this approach in the momentum representation is
the use of  the effective nucleon- nucleon interaction constructed by Love and Franey
 as a two-body $t$-matrix for
the  incident proton scattering on a bound nucleon in the deuteron.
Along with the five-fold cross sections, the proton analyzing power and
the deuteron analyzing powers have been calculated as function of the momentum of
the outgoing fast proton. The results are compared with the data obtained
by the Gatchina-Saclay collaboration.
\end{abstract}

\section{Introduction}

Investigations of the proton-deuteron breakup reaction, or more generally of
nucleon-deuteron reactions, at high collision energies are usually performed with
the aims of studying  the internal deuteron structure, of  searching for non-nucleonic
degrees of freedom, of finding relativistic corrections for the deuteron wave
function (DWF), etc. To a great extent, they have been stimulated by relatively
simple recipes of the plane-wave impulse approximation (PWIA) for evaluation
of the reaction cross sections with unpolarized particles and considering
beam and target asymmetries with polarized protons and deuterons. One has to bear in
mind that the approximation in its most popular form  is equivalent to
the so-called spectator reaction mechanism (with the antisymmetrization
requirement for three-nucleon states not included). It has already been
shown in ref.  \cite {Pun}  that the spectator
approximation, in which one of the nucleons of the deuteron interacts with
the incoming proton and the other one remains largely undisturbed, does not
describe the inclusive ${^1\!H}(d,p)X$ data  on the reaction cross section and the tensor asymmetry $T_{20}$
for the $0^0$-angle proton emission at deuteron kinetic energies  of 1.25 and 2.1 GeV;
this is particularly true  around the kinematic limit where
the dominant reaction channel is $pd \to ppn $. At the same time, according
to refs. \cite {Per,Zb}, the inclusion of more complicated reaction
mechanisms, such as with the double-scattering (DS) and final-state-interaction (FSI)
corrections, to the PWIA results has allowed to improve the theoretical
description.
A similar trend has been observed when explaining the angular space asymmetry
in the deuteron-proton breakup cross sections \cite {Wil,my}
where one can find an approximate estimation of the respective FSI contributions.

Of course, the understanding of the actual role of  two-step reaction
mechanisms depends on the reliability of the additional assumptions that are often made
within the general approach of formal scattering theory, e.g., the Faddeev
scheme for the construction of  transition operators between the initial and
final three-nucleon states (see the review \cite{Gloe96} and refs. therein).
In this context, we would like to note that in ref. \cite {Zb}  the author uses
just the free nucleon propagator
\begin{eqnarray}
\label{g0}
g_0= [E+i0 - K]^{-1},
\nonumber
\end{eqnarray}
with $K$ being the kinetic-energy operator for the three- body system. It enables
one to perform the calculations without significant complications,
due to the neglect of the principal
value part in the decomposition
\begin{eqnarray}
\label{pv}
g_0= P([E - K]^{-1}) - i \pi \delta(E-K),
\nonumber
\end{eqnarray}
but it has also to be taken with considerable doubt.

In the present work we make another endeavour for a proper description of the
 exclusive reaction
$pd \to ppn$ that goes  beyond the scope of the PWIA ansatz whose applicability  has no
firm justification. Therefore, every time, under new kinematic conditions,
one has to take care of a coordinated treatment of the structure and
rescattering effects (in particular, to separate the reaction mechanism from
possible extraordinary deuteron components). Our approach is based on
the Faddeev formulation of the multiple-scattering theory for
the $pd$ system and its full breakup. As in ref. \cite{Zb}, we work with
the nucleon propagator $g_0$. However, unlike ref. \cite{Zb}, we take into account
the dependence of the elementary nucleon-nucleon amplitudes on the internal
motion of a bound nucleon in the deuteron, which is affected by the interaction
with the projectile proton.

Along with the Fermi motion influence some off-shell extrapolations are
introduced for these high-energy vertices by using the effective interaction
from ref. \cite{LF}. Special attention is given to their transformation properties
under  Lorentz boosts (in particular, the Wigner rotations). In addition,
we avoid seemingly artificial truncation in the angular momentum decomposition of
high-energy elementary amplitudes, as made in ref. \cite{Zb}.

We apply the matrix inversion method (MIM) \cite {HaTa,BJ} when calculating
the half-off-energy-shell $t$-matrix that determines the rescattering in
the final (relatively slow) $p-n$ pair. The $p-n$ wave functions in the continuum,
constructed within the MIM, are convenient from an analytical point of view
(cf. the studies  of the exclusive ($e,e'p$) reaction on
the deuteron in refs.\cite{Sch1,Sch2}).
Furthermore the
 spin structure of the  elementary amplitudes is  fully included. This is
of particular significance for the consideration of various polarization
observables.

The paper is organized as follows. The underlying formalism and the leading
order terms in the $NN$ t-matrix for obtaining the $pd \to ppn$ reaction
amplitude are presented in Sec. 2. There one can also find some details concerning
the  application of the MIM. In Sec. 3
we consider the high-energy $NN$ vertices and their transformation from
the NN center-of-mass system to the deuteron rest frame (laboratory system).
The polarization observables in question are given in Sec.4. The results
of our calculations at two kinematics are discussed in Sec.5.
We conclude with Sec. 6. Appendix A contains some useful mathematical relations.

\section{Theoretical Formalism}

The differential cross section of the exclusive reaction
\begin{equation}
\label{r}
p(\vec p)+d(\vec 0)\to p(\vec p_1)+p(\vec p_2)+n(\vec p_3)
\end{equation}
for unpolarized particles in the laboratory system can be written as
\begin{equation}
\frac {d^5\sigma} {dp_1 d\Omega_1 d\Omega_2 }=\frac {(2\pi)^4} {6}
 \frac {E_p} {p}
\frac {p^2_1~  p^2_2~ E_2~ E_3~ |{\cal J}|^2}
{p_2 E_3 +E_2(p_2-p cos \Theta _2 +p_1 cos \Theta _{12})},
\end{equation}
where $\Theta _2$ is the emission angle for the outgoing slow proton and
$\Theta_{12}$ is the angle between the momenta $\vec p_1$ and $\vec p_2$
of the  fast and the slow protons in the final state (see Fig.1).

We introduce the orthonormal basis

\begin{eqnarray}
\label{bas}
\vec z= {\vec p \over |p|}~~,~~~
\vec y=\frac{\vec p\times\vec p_1}{|\vec p\times\vec p_1|}~~,~~~
\vec x=\vec y\times\vec z
\end{eqnarray}
assuming that $\vec p$ and $\vec p_1$ determine the reaction plane so that
$\vec y$ is perpendicular to it.

Following the Alt-Grassberger-Sandhas
 (AGS) formalism of three-body collision theory
(see, for instance, refs. \cite {SZ,Gloc}), the reaction amplitude $\cal J$
is given by the matrix element
\begin{eqnarray}
\label{ampl}
{U}_{pd\to ppn} \equiv \sqrt {2} <123|[1-(1,2)-(1,3)] U_{01}|1(23)>=
\delta (\vec p -\vec p_1-\vec p_2 -\vec p_3){\cal J}
\end{eqnarray}
taken on the energy shell:
\begin{eqnarray}
\label{E}
E=E_p+m_d=E_1+E_2+E_3 .
\end{eqnarray}
Here $E$ is the total energy of the three-nucleon system, $m_d$  the deuteron
mass, and $(i,j)$ the permutation operator for  two nucleons.
The state $|1(23)>$ corresponds to the configuration of nucleons 2 and 3 forming the
deuteron state and nucleon 1 being the projectile, whereas the state
$<123|$ represents the free motion of  three nucleons after the reaction and is the
product of single nucleon states.

Iterating the AGS equations for the  rearrangement operators, one obtains
the transition operator $U_{01}$ as
\begin{eqnarray}
\label{next}
U_{01}=g_0^{-1}+t_2+t_3+t_1 g_0t_2 +t_1 g_0t_3+t_2 g_0t_3+t_3 g_0t_2+O(t^3) .
\end{eqnarray}
By definition, $t_3=t_{12}$ is a two-nucleon transition operator
(the other can be obtained via  cyclic permutations).
Obviously, the  term $g_0^{-1}$ does not contribute to
the on-shell amplitude (\ref {ampl}).

We are interested in such  kinematic situations where, firstly,
one of the final protons
carries away a considerable portion of the initial momentum $\vec p$ and,
secondly, the momentum $q=|\vec p -\vec {p_1}|$, transferred to the p-n pair,
is small in comparison with the nucleon mass $m$.
It enables us to use various nonrelativistic models for the DWF.
Under such conditions we retain the first- and second- order contributions to
the multiple-scattering expansion (\ref {next}), neglecting the corrections
due to the so-called recoil reaction mechanism for which the high-momentum
proton leaves the deuteron without direct knock-out.
In addition, small contributions from the DS with participation of a fast
nucleon, i.e. the $t(fast) g_0 t(fast)$ terms, are disregarded,
what is in accordance with the conclusions in ref. \cite {Zb}.
After this, the matrix element $ U_{pd \to ppn}$
can be approximated by
\begin{eqnarray}
\label{am}
U_{pd \to ppn}&=&\sqrt {2} <123|[1-(2,3)][1+t_{23} g_0]t_{12}^{sym}|1(23)>
\end{eqnarray}
with the symmetrized NN t-operator $t_{12}^{sym}=[1-(1,2)]t_{12}$.

The three-body operator $\omega_{23} \equiv 1+t_{23} g_0$ acting on the state
$<123|$ yields
\begin{eqnarray}
\label{star}
<123|\omega_{23}&=&\lim _{\varepsilon\to 0+}
<123|\frac {i\varepsilon}{E + i\varepsilon -K}
\left [1+t_{23}\frac {1}{E + i\varepsilon-K} \right ]
\\
&=&
\lim _{\varepsilon\to 0+}<123|\frac {i\varepsilon}{E + i\varepsilon-K-V_{23}},
\nonumber
\end{eqnarray}
where $V_{23}$ is the potential operator of the two-nucleon interaction
\footnote{A possible way for the construction of a NN interaction starting
from a fundamental Hamiltonian of interacting meson and nucleon fields has
been shown recently in ref.  \cite {SS}}. Here we have employed the relation
\begin{eqnarray}
\label{hilbert}
\frac {1}{E + i\varepsilon-K-V_{23}} = \frac {1}{E + i\varepsilon-K} +
\frac {1}{E + i\varepsilon-K} t_{23} \frac {1}{E + i\varepsilon-K}
\end{eqnarray}
for the three-body operator $t_{23}$, which obeys a Lippmann-Schwinger (LS)
equation with $V_{23}$ as  driving term.

Further, since $K=K_1 + K_2 + K_3$ and, by convention, $K_i |123> = E_i |123>$,
 one can rewrite
Eq. (\ref {star}) in the form
\begin{eqnarray}
\label{om1}
<123|\omega_{23} =
\lim _{\varepsilon\to 0+}<123|\frac {i\varepsilon}
{E-E_1+i\varepsilon-K_2-K_3-V_{23}}
\end{eqnarray}
or
\begin{eqnarray}
\label{om2}
<123|\omega_{23} = <123| \left[ 1 + t_{23}(E-E_1) g_{23}(E-E_1) \right ] .
\end{eqnarray}
The operator $g_{23}(E-E_1) \equiv (E-E_1+i0-K_2-K_3)^{-1} $ is a free
propagator for the (23)-subsystem, and the scattering operator
$t_{23} (E-E_1)$ satisfies the LS equation
\begin{eqnarray}
\label{LS}
t_{23}(E-E_1) = V_{23} + V_{23} g_{23}(E-E_1) t_{23}(E-E_1) .
\end{eqnarray}

Let us rewrite the matrix element (\ref{am}) indicating explicitly
the particle quantum numbers,
\begin{eqnarray}
U_{pd\to ppn}=\sqrt {2}
<\vec {p_1} \mu_1 \tau_1,\vec {p_2} \mu_2 \tau_2,\vec {p_3} \mu_3 \tau_3|
[1-(2,3)] \omega_{23} t^{sym}_{12} |\vec {p} \mu \tau ,\psi _{1 M_d 0 0} (23)>,
\nonumber
\end{eqnarray}
with the spin and isospin  projections denoted as
$\mu$ and $\tau$ respectively. We assume that $\tau = \tau_1 =\tau_2 = 1/2 $ and
$\tau_3 = - 1/2 $.
Inserting the unity operator
\begin{eqnarray}
{\bf 1}=\int d\vec p^\prime
|\vec p^\prime {\mu}^\prime\tau ^\prime>
<\vec p^\prime {\mu}^\prime\tau ^\prime|,
\nonumber
\end{eqnarray}
we arrive at the following expression for the reaction amplitude
\begin{eqnarray}
\label {1}
{\cal J}&=&\sum _{T T^\prime} C_{TT^\prime}
<{1\over 2} \mu_2 {1\over 2} \mu_3|S M_S>
 <{1\over 2} {\mu}^\prime _2 {1\over 2} {\mu}^\prime _3|S M^\prime _S>
\nonumber\\
&&\int d\vec p _r {^\prime }
\left<\vec p_r, S M_S\left|1 + m \frac{t^{ST}(E-E_1)}
{\vec {p_r} ^2 -\vec {p_r}^{\prime 2}+i0 }\right|
\vec p _r {^\prime }, S M^\prime _S \right>
\nonumber\\
&&<\vec {p_1} \mu_1,(\vec {p_r}^\prime + \vec q/2)~ {\mu}^\prime _2|
t^{T^\prime}_{sym}(E - E_3^\prime)
|\vec p \mu, (\vec {p_r}^\prime -\vec q/2)~ {\mu}^{\prime \prime}>
\\
&&<{\mu}^{\prime\prime} {\mu}^\prime_3|\psi _{1M_d}(\vec p_r {^\prime } -\vec q/2)>
~-~
(2\leftrightarrow 3),
\nonumber
\end{eqnarray}
where $E_3^\prime=\sqrt{m_N^2+(\vec p_r{^\prime} -\vec q/2)^2}$. Here we have 
introduced the  momentum transfer
$\vec q=\vec p -\vec p_1$,  relative momenta
$\vec p_r={1\over 2}(\vec p_2 -\vec p_3)$ and
$\vec {p_r}^\prime ={1\over 2}(\vec {p_2}^\prime  -\vec {p_3}^\prime)$,
and  isotopic coefficients
\begin{eqnarray}
C_{TT^\prime}= {1\over 2}
\left[ {{(-1)^T}\over 2} \delta _{T^\prime 0} +
\left( 1+ {{(-1)^T}\over 2}\right) \delta _{T^\prime 1} \right].
\end{eqnarray}
Henceforth, all summations over dummy discrete indices are implied.

In momentum representation the DWF  $\psi _{1 M_d}(\vec k)$  with
 spin projection $M_d$ is written as
\begin{eqnarray}
|\psi _{1 M_d}(\vec k)>=\sum _{L=0,2}\sum_{M_L=-L}^{L}
<L M_L 1 M_s|1 M_d> u_L (k) Y_L^{M_L}(\hat k) | 1 M_s>,
\end{eqnarray}
with the spherical harmonics $Y_L^{M_L}(\hat k)$ and the Clebsh-Gordon
coefficions in standard form.
In our calculations, we have employed the following parametrizations of
the S- and D- state wave functions
\begin{eqnarray}
u_0 (p)=\sqrt{{2\over\pi}}\sum _{i}\frac {C_i}{\alpha_i ^2 +p^2}~~,~~
u_2 (p)=\sqrt{{2\over\pi}}\sum _{i}\frac {D_i}{\beta_i ^2 +p^2}
\end{eqnarray}
as proposed in refs. \cite {NN,B} .

The wave function of the final p-n pair
\begin{eqnarray}
<\psi^{(-)}_{\vec p_r S M_S T M_T}|\vec p _r {^\prime } S M^\prime _S T M_T>
=
\delta (\vec {p_r} - \vec {p_r}^\prime)\delta _{M_s M_s^\prime}+
\nonumber\\
\frac{m_N}{\vec {p_r} ^2 -\vec {p_r}^{\prime 2}+i0 }
< \vec p_r S M_S | t^{ST}| \vec p _r {^\prime } S M^\prime _S >
\end{eqnarray}
contains the FSI part, which can be taken  in different ways.
First of all, we note that in ref.  \cite {Wil} the t-matrix
elements were obtained by solving Eq. (\ref {LS}) with a separable NN
interaction.
Of course, these solutions are not able to reproduce the off-shell behaviour
of the NN t-matrix with more realistic two-body forces.
Another approximation relies on the  neglect of the principal value
part in
\begin{eqnarray}
{1\over {\vec p_r^2-\vec p^{\prime 2}_r+i0}}=
-i\pi\delta (\vec p_r^2-\vec p^{\prime 2} _r)+P\left[ {1\over
 {\vec p_r^2-\vec p^{\prime 2}_r}}\right] .
\end{eqnarray}
Then, the FSI term can be expressed through the on-energy-shell
t-matrix, i.e. the  respective phase shifts. Third procedure ( see, ref.
\cite {Bel} and refs. therein) uses some artificial off-shell extrapolation of
the t-matrix with energy-dependent prescriptions for the introduction of a
proper form factor.

In this paper we use the MIM
applied to study  the deuteron  electrodisintegration \cite {Sch1,Sch2}.
As in ref. \cite {Sch1}, we consider the truncated partial-wave expansion,
\begin{eqnarray}
&&<\psi^{(-)}_{\vec p_r S M_S T M_T}|\vec p_r {^\prime } S M^\prime _S T M_T>=
\delta _{M_S M^\prime _S} \delta (\vec p_r -\vec p_r {^\prime})+
\\
&&\sum _{J=0}^{J_{max}} \sum _{M_J=-J}^{J}
Y_l^m (\hat p_r) <l m S M_S|J M_J>
\psi _{l l^\prime} ^\alpha (p_r {^\prime})
<l^\prime m ^\prime S M_S ^\prime|J M_J>
{Y} _{l^\prime}^{\ast m^\prime} (\hat p_r{^\prime}),
\nonumber
\end{eqnarray}
where $J_{max}$ is the maximum value of the total angular momentum in
n-p partial waves and $\alpha =\{ J,S,T\}$ is the set of conserved quantum
numbers.
The radial  functions $\psi^\alpha _{ll^\prime}(p_r{^\prime})$ are related via
\begin{eqnarray}
\label {2}
\psi _{l l^\prime} ^{JST} (p_r {^\prime})=\sum _{l^{\prime\prime}}
O_{l l^{\prime\prime}}\varphi _{l^{\prime\prime}l^\prime}(p_r^{\prime})-
\frac {\delta (p_r ^{\prime} -p_r)}{p_r^2} \delta _{l l^\prime}
\end{eqnarray}
to the partial wave functions
$\varphi_{l^{\prime\prime}l^\prime}^\alpha (p_r^\prime)$, which
have the asymptotics of  standing waves. The coefficients
$O_{ll^{\prime\prime}}$ can be expressed in  terms of the
corresponding phase shifts and mixing parameters \cite {Sch1}.

Within the MIM, the functions $\varphi_{ll^\prime}^\alpha (p_r^\prime)$
can be represented as
\begin{eqnarray}
\varphi _{ll^\prime} ^\alpha (p_r {^\prime})=\sum _{j=1}^{N+1}
B_{ll^\prime} ^\alpha (j) \frac {\delta (p_r ^{\prime} -p_j)}{p_j^2},
\end{eqnarray}
where the coefficients $B_{ll^\prime} ^\alpha (j)$ fulfill a  set of
 linear algebraic equations approximately equivalent to the
$LS$ integral equation for the $n-p$ scattering problem.
Here N is the dimension of this set,
 $p_j$ are the grid points associated with the Gaussian nodes over the
interval [-1,1] and $p_{N+1}=p_0$
(details can be found in ref. \cite {KFTI}).
It should be noted that in this way the nucleon wave function is 
expressed by a series of
 $\delta$-functions allowing one to reduce a triple integral
 in Eq. (\ref 1) to a double one. In addition, the method offers the opportunity to
 consider the nucleon wave function in the continuum directly in momentum
space what simplifies all the calculations a lot.

\section{High-Energy Vertex and Its Transformations}

As seen from Eq. (13), we have to deal with the high-energy matrix elements
\begin{eqnarray}
\label{31}
t^{T^\prime} (E_{on-shell} ;\vec p _1 {\mu}^\prime _1 ,
\vec p^\prime _2 {\mu}^\prime _2 ,\vec p \mu _1, \vec p ^\prime \mu _2  ) \equiv
\nonumber\\
<\vec p _1 {\mu}^\prime _1 , \vec p^\prime _2 {\mu}^\prime _2|t^{T^\prime} (E_{on-shell}) |
\vec p \mu _1, \vec p ^\prime \mu _2 > ,
\end{eqnarray}
where $p_1\sim p \geq m$ and $E_{on-shell} \geq 2m$ , which are determined  not only
on the energy shell
\begin{eqnarray}
\label{32}
E_{\vec p _1} + E_{\vec p^\prime _2} =E_{on-shell}= E_{\vec p } + E_{\vec p ^\prime }
\end{eqnarray}
but largely beyond it for N-N scattering with arbitrary values of the  total momentum
$\vec p + \vec p ~^\prime = \vec p _1 + \vec p_2 ~^\prime $ (see Fig. 1). 
The latter is conserved owing to
the requirement of translational invariance.

The frame dependence of these off-shell quantities, in particular, their expression in
terms of the $t$-matrix elements in the N-N center-of-mass system
(c.m.s.) can be found making use of some
prescriptions \cite{Hell76,Garc77} of the relativistic potential theory (RPT) for relating
a two-body off-energy-shell $t$-matrix between different frames of reference.
First, let us recall the equation
\begin{eqnarray}
\label{33}
t(E;\vec k^\prime _1 {\mu}^\prime _1 , \vec k^\prime _2 {\mu}^\prime _2 ,
\vec k_1 \mu _1, \vec k_2 \mu _2  )&&=N^\prime N
<{\mu}^\prime _1|W^\dagger(\vec u^\prime, \vec k^\prime _1)|\nu^\prime _1>
\\
&&<{\mu}^\prime _2|W^\dagger(\vec u^\prime, \vec k^\prime _2)|\nu^\prime _2>
t(E,\vec K; \vec k^{\prime {*}} {\nu}^\prime _1 {\nu}^\prime _2; \vec k^{*} \nu_1,\nu_2 )
\nonumber
\\
&&<{\nu}_1|W(\vec u, \vec k_1)|\mu_1> <{\nu}_2|W(\vec u, \vec k_2)|\mu_2> ,
\nonumber
\end{eqnarray}
which connects the on-momentum-shell matrix elements in  two bases\footnote{Henceforth,
the asterisk is used to mark  quantities in the c.m.s.}. One of them,
$\{| \vec k_1 \mu _1, \vec k_2 \mu _2 > \} $, is a direct product
$\{| \vec k_1 \mu _1>\otimes|\vec k_2 \mu _2 > \}$ where the single-nucleon states
$| \vec k_a \mu _a>$ satisfy the following orthonormality and completeness conditions
\begin{eqnarray}
\label{34}
<\vec k^\prime _a \mu^\prime _a | \vec k_a \mu _a> =
\delta(\vec k^\prime _a - \vec k_a) \delta_{\mu^\prime _a,\mu _a }
\end{eqnarray}
and
\begin{eqnarray}
\label{35}
\int d\vec k_a |\vec k_a \mu _a><\vec k_a \mu _a| = 1 .
 \end{eqnarray}
By assumption, they are related to the states $|[m,\frac12], \vec k, \mu> =
\sqrt{2 E_{\vec k}} |\vec k \mu>$ of the so-called canonical basis, which is
 well known
in  elementary particle physics (see, e.g., Ch. IV  of ref. \cite{Gas66}). 
The latter is
transformed by an irreducible representation of the Lorentz group.
In particular, we
have under the Lorentz boost $L(\vec v)$ corresponding to the motion of 
the reference
frame with  velocity $\vec v$,
\begin{eqnarray}
\label{36}
|[m,\frac12], \vec k_a, \mu_a>^{tr} = |[m,\frac12], {\vec k}^{tr}_a, \nu_a >
<\nu_a|W(\vec u, \vec k_a)|\mu_a>  \qquad (a=1,2),
\end{eqnarray}
where
\begin{eqnarray}
\label{37}
W(\vec u, \vec k_a) = \exp (- \imath \vec \sigma_a\cdot \vec \omega_a ) =
\nonumber\\
\cos \frac{\omega_a}{2}
\left\{ 1 - \imath\frac{\vec\sigma_a\cdot[\vec u\times \vec k_a]}{(1+u_0)m +
E^{tr}_a + E_a} \right \}
\end{eqnarray}
is the Wigner rotation operator in the spin space of the a-th nucleon. 
Here we 
introduce the four-velocity 
$u=(u_0, \vec u)=(u_0, u_0 \vec v)$ with a Lorentz scalar
$u^2=1$ and denote 
the nucleon energies in the initial
and moving frames of reference
by $E_a=\sqrt{{\vec k_a }^2 + m^2}$ and
$E^{tr}_a = \sqrt{{\vec k}^{tr 2}_a + m^2} $. 
The angle $\omega_a$ of this rotation about the axis
$[\vec u\times \vec k_a]$ is determined by the relation
\begin{eqnarray}
\label{38}
\tan \frac{\omega_a}{2} =
\frac{|[\vec u\times \vec k_a]|}{(1+u_0)m + E^{tr}_a + E_a}
\end{eqnarray}
(see, for instance, Eq. (32) in ref. \cite{Ritus} or Eq. (C.7) in ref. \cite{Sch2} ).
 In formulae
(\ref{36})-(\ref{38}) 
the symbol ${}^{tr} $
belongs to the transformed quantities, so that the transformed four-momentum is
\begin{eqnarray}
\label{39}
k^{tr}_a = (E^{tr}_a ,{\vec k}^{tr}_a  ) = L(\vec v) k_a = L(\vec v) (E_a , \vec k_a) .
\end{eqnarray}

Since we are interested in a specific transition to the N-N c.m.s. simultaneously
for the initial and final N-N states, all we need is to put in Eq. (\ref{33})
$\vec u = \vec K/\sqrt{s}$ and $\vec u^\prime = \vec K^\prime /\sqrt{s^\prime}$, where
$\vec K = \vec k_1 + \vec k_2 = \vec k^\prime_1 + \vec k^\prime _2=\vec K^\prime $ is
the total
momentum of the initial(final) N-N pair. Normally, the Lorentz scalars $s$ and
$s^\prime$ are defined as $s=(k_1 + k_2)^2= E^{* 2}$ and
$s^\prime= ( k^\prime_1 + k^\prime _2 )^2 = E^{* \prime 2} $.

The corresponding basis $\{|\vec K \vec k^{*} \mu_1 \mu_2> \}$ is composed of
 two-nucleon states with relative momentum $\vec k^{*}$ between  two nucleons,
measured in their c.m.s., and  total momentum $\vec K$. In accordance with
the customary recipes (see, for instance, Sect. 28 in ref. \cite{Werle}), every state
$|\vec K \vec k^{*} \mu_1 \mu_2>$ can be represented as a product state
$|[m,\frac12], \vec k^{*}, \mu_1> |[m,\frac12], - \vec k^{*}, \mu_2>$ which is subject
to a subsequent Lorentz boost with  velocity $ \vec u = - \vec K/E^{*} $. Doing so,
we get
\begin{eqnarray}
\label{310}
|\vec K \vec k^{*} \mu_1 \mu_2> = N | \vec k_1 \nu _1>|\vec k_2 \nu _2 >
<\nu_1|W^\dagger(\vec K/E^{*}, \vec k_1)|\mu _1>
\nonumber\\
<\nu_2|W^\dagger(\vec K/E^{*}, \vec k_2)|\mu _2> .
\end{eqnarray}
We have used the property $W(- \vec u, \vec k) = W^\dagger(\vec u, \vec k) $ and
the relations $W(\vec K/E^{*} , \vec k^{*})= W(\vec K/E^{*}, \vec k_1) $ and
$W(\vec K/E^{*} ,- \vec k^{*}) = W(\vec K/E^{*} , \vec k_2) $ .
In this context, note that for given momenta $\vec k_1$  and $\vec k_2$ in an arbitrary
frame of reference
\begin{eqnarray}
\label{311}
\vec k^{*} = \vec k^{*}_1 = - \vec k^{*}_2 = \frac{E_2 + E^{*}_2}{E_1 + E_2 + E^{*}} \vec k_1-
\frac{E_1 + E^{*}_1}{E_1 + E_2 + E^{*}} \vec k_2 ,
\end{eqnarray}
where $E^{*} = E^{*}_1 + E^{*}_2$ is the respective invariant mass (the total energy
in the c.m.s.).

In order to provide the orthonormality and completeness relations
\begin{eqnarray}
\label{312}
<\vec K^\prime \vec k^{\prime *} \mu^\prime_1 \mu^\prime_2 | \vec K \vec k^{*} \mu_1 \mu_2> =
\delta(\vec K^\prime - \vec K) \delta (\vec k^{\prime *} - \vec k^{*} )
\delta_{\mu^\prime_1,\mu_1} \delta_{\mu^\prime_2,\mu_2 }
\end{eqnarray}
and
\begin{eqnarray}
\label{313}
\int d\vec K d\vec k^{*} | \vec K \vec k^{*} \mu_1 \mu_2><\vec K \vec k^{*} \mu_1 \mu_2| = 1
\end{eqnarray}
the normalization factor $N$ in Eq. (\ref{310}) (see also Eq. (\ref{33})) should be
\begin{eqnarray}
\label{314}
N = \left[ J (\vec k_1, \vec k_2 ) \right]^{ - \frac12},
\end{eqnarray}
where $J (\vec k_1, \vec k_2 )$ is the  Jacobian for the transformation between
the $\vec k_1 \vec k_2 $ -- and $ \vec K \vec k^{*} $ -- representations, i.e.
\begin{eqnarray}
\label{315}
J (\vec k_1, \vec k_2 ) = \frac{ E^{*}_1 + E^{*}_2}{E_1 + E_2}
\frac{E_1 E_2}{E^{*}_1 E^{*}_2} .
\end{eqnarray}
Analogously, one has
\begin{eqnarray}
\label{316}
J (\vec k^\prime_1, \vec k^\prime_2 ) =
\frac{ E^{\prime *}_1 + E^{\prime *}_2}{E^\prime_1 + E^\prime_2}
\frac{E^\prime_1 E^\prime_2}{ E^{\prime *}_1 E^{\prime *}_2} ,
\end{eqnarray}
and the normalization factor $N^\prime$ for the final N-N pair is
\begin{eqnarray}
\label{317}
N^\prime = \left[ J (\vec k^\prime_1, \vec k^\prime_2 ) \right]^{ - \frac12} .
\end{eqnarray}

The next result \cite{Garc77}
\begin{eqnarray}
\label{318}
t(E;\vec K; \vec k^{\prime {*}} {\nu}^\prime _1 {\nu}^\prime _2; \vec k^{*} \nu_1,\nu_2 ) =
F t(\sqrt{E^2 - {\vec K}^2};\vec k^{\prime {*}} {\nu}^\prime _1 {\nu}^\prime _2;
\vec k^{*} \nu_1,\nu_2 ) ,
\end{eqnarray}
with
\begin{eqnarray}
\label{319}
F = \frac{E}{\sqrt{E^2 - {\vec K}^2}}
\frac{\sqrt{E^2-{\vec K}^2}+\sqrt{(E^\prime_1+E^\prime_2)^2-{\vec K}^2}}{E+E^\prime_1+E^\prime_2}\cdot
\nonumber\\
\frac{\sqrt{E^2-{\vec K}^2}+\sqrt{(E_1+E_2)^2-{\vec K}^2}}{E+E_1+E_2},
\end{eqnarray}
enables one to express the fully-off-shell matrix elements with nonzero total momentum
through the others with zero total momentum. The latter are the fully-off-shell $t$-matrix
elements in the c.m.s..

Using Eqs. (\ref{33}) and (\ref{318}) in our evaluation of the matrix elements
(\ref{31}), we approximate them with the ones in the N-N laboratory system (LAB) 
with $\vec p ~^\prime = 0$.
This appears to be reasonable since 
$\vec p ~^\prime = \vec {p_r}^\prime -\vec q/2$ (see Eq. (\ref{1}))
is an internal nucleon momentum in the deuteron. In addition, we neglect the deuteron
binding energy, and thus
\begin{eqnarray}
\nonumber\\
E_{on-shell} = E_{\vec p} + m \equiv E ,~~~ E^\prime = E^\prime_1+E^\prime_2 = E_{\vec p_1} + E_{\vec p_2^\prime},
\nonumber\\
s = 2mE,\qquad
s^\prime ={E^\prime }^2-\vec p ^2 .
\nonumber
\end{eqnarray}
Of course, in order to avoid any confusion, one should distinguish 
between the total two-body
energy $E$ in this section and the collision energy in Eq. (\ref{E}).

Then the transformation of interest for this half-off-shell $t$-matrix can be written
in the compact form,
\begin{eqnarray}
\label{320}
t^{LAB} \equiv t^{T^\prime} (E;\vec p _1 {\mu}^\prime _1,\vec q {\mu}^\prime _2 ,\vec p \mu _1
, \vec 0 \mu _2  ) =
\Phi W^\dagger(\vec u^\prime, \vec p_1) W^\dagger(\vec u^\prime, \vec q) t^{CM}
\end{eqnarray}
with
\begin{eqnarray}
\label{321}
\Phi = \frac{1}{4}\sqrt{\frac{ss^\prime}{m E_{\vec p_2^\prime} E_{\vec p} E_{\vec p_1}}}
\sqrt{\frac{E E^\prime}{\sqrt{ss^\prime}}} \frac{\sqrt{s} + \sqrt{s^\prime}}{E + E^\prime }
\end{eqnarray}
or
\begin{eqnarray}
\label{322}
\Phi = \sqrt{\left(1-\frac{T_{\vec p}}{2E_{\vec p}}\right)
\left(1-\frac{T_{\vec p_1}}{2E_{\vec p_1}}\right)}
\sqrt{\frac{E E^\prime}{4m E_{\vec p_2^\prime}}} \frac{\sqrt{s} + \sqrt{s^\prime}}{E + E^\prime },
\end{eqnarray}
where $T_{\vec p}$ $(T_{\vec p_1})$ is the kinetic energy of the incident (outgoing) proton,
and $t^{CM} \equiv t(\sqrt{s};\vec k^{\prime {*}}; \vec k^{*} )$ is the half-off-shell
$t$-matrix in the N-N c.m.s. Further, a simple calculation yields
\begin{eqnarray}
\label{323}
W^\dagger(\vec u^\prime, \vec p_1) = \cos\frac{\omega_1}{2}
[1 + i \vec \sigma_1\cdot\vec y \tan\frac{\omega_1}{2}] ,
\end{eqnarray}
and
\begin{eqnarray}
\label{324}
W^\dagger(\vec u^\prime, \vec q) = \cos\frac{\omega_2}{2}
[1 - i \vec \sigma_2\cdot\vec y \tan\frac{\omega_2}{2}] ,
\end{eqnarray}
with
\begin{eqnarray}
\label{325}
\tan\frac{\omega_1}{2} = \frac{p}{\sqrt{s^\prime}} \frac{p_1}{K_1} \sin \theta_1,
\end{eqnarray}
and
\begin{eqnarray}
\label{326}
\tan\frac{\omega_2}{2} = \frac{p}{\sqrt{s^\prime}} \frac{p_1}{K_2} \sin \theta_1,
\end{eqnarray}
where $\theta_1$ is the angle between the momenta
$\vec p$ and $\vec p_1$, i.e. the emission angle of the fast
proton in the deuteron rest frame. Explicit expressions for the coefficients
$K_1$ and $K_2$ are given in the Appendix.
From Eqs. (\ref{323}) and (\ref{324})
 we see that the Wigner operators  correspond to rotations
in opposite directions about the $\vec y$ - axis normal to the reaction plane.

When evaluating the half-off-shell elements $t(\sqrt{s};\vec k^{\prime {*}}; \vec k^{*} )$
in Eq. (\ref{320}) we have started with the Love-Franey model
used to construct 
the on-shell N-N $t$-matrix in each N-N channel (triplet-odd, triplet-even, etc.).
According to ref. \cite{LF} (see Eq. (12) in the first paper), 
the corresponding matrix elements
are expressed through the effective N-N interaction operators sandwiched  between
the initial and final plane-wave states. We extend this construction to the off-shell
case using the same operators with  complex strengths depending  on $\sqrt{s}$
and allowing the initial and final momenta to assume the current values of $\vec k^{*} $
and $\vec k^{\prime {*}}$ . Obviously, such an off-shell extrapolation does not change
the general spin structure
\begin{eqnarray}
\label{51}
t_{NN}^{CM}&=&A+B(\vec\sigma_1 \hat n^*)(\vec\sigma_2 \hat n^*)+
C(\vec\sigma_1 +\vec\sigma_2 )\cdot \hat n^* +
\nonumber\\
&&D(\vec\sigma_1 \hat q^*)(\vec\sigma_2 \hat q^*) +
F(\vec\sigma_1 \hat Q^*)(\vec\sigma_2 \hat Q^*),
\nonumber
\end{eqnarray}
where
\begin{equation}
\hat q^*=\frac {\vec k^*-\vec k^{*\prime}}{|\vec k^* -\vec k^{*\prime}|}~~,~~
\hat Q^*=\frac {\vec k^*+\vec k^{*\prime}}{|\vec k^* +\vec k^{*\prime}|}~~,~~
\hat n^*=\frac {\vec k^* \times \vec k^{*\prime}}{|\vec k^*
\times\vec k^{*\prime}|},
\end{equation}
inherent in the on-shell N-N $t$-matrix in c.m.s. (see, for instance, ref. \cite {GW}).
Of course, this structure is subject to certain changes under the transformation
(\ref{320}) so that
\begin{eqnarray}
\label{52}
t_{NN}^{lab}=\cos\frac{\omega_1}{2}  \cos\frac{\omega_2}{2} \Phi
\{F_1+F_2(\vec\sigma_1 \vec y)+F_3(\vec\sigma_2 \vec y)+
F_4(\vec\sigma_1 \vec x)(\vec\sigma_2 \vec x)+
\nonumber\\
F_5(\vec\sigma_1 \vec y)(\vec\sigma_2 \vec y)+
F_6(\vec\sigma_1 \vec z)(\vec\sigma_2 \vec z)+
F_7(\vec\sigma_1 \vec x)(\vec\sigma_2 \vec z)+
F_8(\vec\sigma_1 \vec z)(\vec\sigma_2 \vec x)\} .
\end{eqnarray}
Relationships between the structure functions $F_i$ in the laboratory and $A$ to $F$ in 
the center- of -mass systems are given
in the Appendix.

\section{ Polarization Observables for Proton-Deuteron Breakup.}

The  reaction amplitude ${\cal J}$ resulting from our calculation, allows us to find
the spin polarization coefficients
\begin{eqnarray}
\label {41}
C _{i,j,k,l,n}={1\over C_0} Tr\left\{ {\cal J} {\cal P}_i \sigma_j {\cal J}^\dagger \sigma_k
\sigma_l \sigma_n \right\}
\end{eqnarray}
with
\begin{eqnarray}
\label {42}
C_0 = Tr \left\{ {\cal J} {\cal J}^\dagger \right \} ,
\end{eqnarray}
where the first and second indices i and j refer to 
the deuteron
and the incident proton
 polarizations, respectively, and the remaining ones specify the  nucleon
polarizations in the final state. The standard operators ${\cal P}_i (i=0,1,...,9)$
introduced by Goldfarb (see, e.g., Eqs. (2.17) in ref. \cite{Ohl72}) form the basis 
for
the $3 \times 3$ spin 1 space. 
If a particle in the entrance channel is unpolarized
or the polarization of a final nucleon is not analyzed, the corresponding index is
assumed equal to zero.

In this paper, we consider the first- order observables related to the polarization of
 initial particles for coplanar geometry. In this context,
let us determine the analyzing power of the projectile proton
\begin{eqnarray}
\label{43}
C_{0,y,0,0,0}={2\over C_0}{\it Im}
\sum <m_1,m_2,m_3|{\cal J}|{1\over 2} M_d><M_d,-{1\over 2}|{\cal J}|m_1,m_2,m_3>^* ,
\end{eqnarray}
and the vector and tensor analyzing powers of the deuteron
\begin{eqnarray}
\label{44}
C_{y,0,0,0,0}&=&{\sqrt {2}\over C_0}{\it Im}
\sum\{<m_1,m_2,m_3|{\cal J}|m 1><0 m|{\cal J}|m_1,m_2,m_3>^* -
\nonumber\\
&&<m_1,m_2,m_3|{\cal J}|m -1><0 m|{\cal J}|m_1,m_2,m_3>^*\},
\nonumber\\
C_{yy,0,0,0,0}&=&-{1\over {2 C_0}}
\sum\{|<m_1,m_2,m_3|{\cal J}|m 1>|^2 + |<m_1,m_2,m_3|{\cal J}|m -1>|^2 -
\nonumber\\
&&
2|<m_1,m_2,m_3|{\cal J}|m 0>|^2 +
\\
&&6{\it Re}<m_1,m_2,m_3|{\cal J}|m 1><-1 m|{\cal J}|m_1,m_2,m_3>^*\}
\nonumber\\
C_{zz,0,0,0,0}&=&{1\over  C_0}
\sum\{|<m_1,m_2,m_3|{\cal J}|m 1>|^2 + |<m_1,m_2,m_3|{\cal J}|m -1>|^2 -
\nonumber\\
&&
2|<m_1,m_2,m_3|{\cal J}|m 0>|^2 \} .
\nonumber
\end{eqnarray}
Of course, the symbol ${\sum }$ means summation over the particle
spin projections. Here, unlike in Eq. (\ref{1}) (more exactly, in its l.h.s.),  these
projections are explicitly indicated. As a consequence of parity conservation, the vector
analyzing powers of the
deuteron $C_{x,0,0,0,0}$, $C_{z,0,0,0,0}$ and the proton $C_{0,x,0,0,0}$, $C_{0,z,0,0,0}$
are equal to zero in coplanar geometry. Moreover, the tensor analyzing power $C_{xx,0,0,0,0}$
is not independent owing to the  relation 
$C_{xx,0,0,0,0}+C_{yy,0,0,0,0}+C_{zz,0,0,0,0}=0$
(cf. Eq. (2.18) of ref. \cite{Ohl72}). Thus, we have only the above four independent
first- order observables.

\section{Results and Discussion}

Using  expression (\ref{1}) for the reaction amplitude, the exclusive cross sections
and the first-order polarization observables have been calculated at a 
 projectile energy of 1 GeV in the deuteron rest frame.
All the calculations  have been performed with the Paris potential
\cite{NN}.
  We have confined ourselves to the
coplanar geometry (with  momentum $\vec p_2$ lying in the reaction plane) and
considered  two kinematics. 

Kinematics I is equivalent to the Gatchina
experiment \cite {Bel,Gr} as the $pp$ coincidence cross sections have been measured on
the left  of the quasi-free peak (QFP), far from its maximum (Fig. 2). Recall
that  the latter corresponds to  proton
elastic scattering on a free proton at rest
 in case of a  spectator momentum $p_3 = 0$. 
 In fact, the four-momentum conservation
for  reaction (\ref{r}) in the deuteron rest 
frame yields  the Bjorken variable
$x=q^2_{\mu}/(2m\omega) = 1$, where $q^2_{\mu} = {\vec q}^2 - {\omega}^2 = - t $.
Of course, the equality $x=1$ is valid only up to  small corrections
due to binding effects in the deuteron.
The FSI distortions have been taken into account in $p-n$ partial scattering states with
$J_{max}$=2,3, and 4. We see that the results 
obtained for the case with $J_{max}=3 $ are practically converged and may be
assigned a
reasonable predictive power (cf. Figs. 2-6).

The experimental conditions have been chosen so that FSI contributions could
be neglected. Indeed, the FSI give insignificant
corrections for the differential cross section with respect to the
PWIA predictions (Fig. 2). However, 
their effects are more pronounced
for the polarization observables. 
In this connection, especially the tensor analyzing
power of the deuteron $C_{yy,0,0,0,0}$ (Fig. 5) and the vector analyzing power
of the proton $C_{0,y,0,0,0}$ (Fig. 3) are of great interest.
Note in particular that the PWIA curve for $C_{yy,0,0,0,0}$
(Fig.5) has a very quickly changing behaviour and comes from a
negative region up to nearly the upper limit of 1 while the
full result lies only in the positive region, has a modest
slope and varies from $\sim$ 0.5 to 0.8.
We observe a different behaviour  for the other tensor analyzing power
of the deuteron  $C_{zz,0,0,0,0}$ (Fig. 6). One can see that this observable 
 in PWIA      
assumes small negative values while the calculations taking account of the
FSI produce a positive result with very quickly changing values.
In addition, the result
obtained with $J_{max}=2$ is significantly different
from the one with $J_{max}=3$, what is not so evident for the other
observables.
For  the vector analyzing powers of the proton $C_{0,y,0,0,0}$ (Fig. 3) and the
deuteron  $C_{y,0,0,0,0}$ (Fig. 4)
the PWIA and FSI curves have a similar behaviour, but
their absolute values are  about two to three times different.
Thus, we can conclude that the polarization observables are very sensitive
to reaction mechanisms. Unfortunately, only the differential cross section
has  been measured
 under such kinematic conditions.

The same observables in kinematics II are presented in
Figs. 7-11. The projectile proton energy is 1 GeV as in the previous case.
The scattering angles of the  fast and slow protons are
$\theta_1=6.77^0$ and $\theta_2=83^0$, respectively.
In such kinematics  we have more favourable conditions for the
applicability of our approach. Really, while  the $q$ $(p_3)$ - values for
kinematics I
range from 500 (300) to 650 (570) MeV/c, the momentum transfer $q$ 
and the neutron momentum $p_3$
 for the kinematics II with a cut at
$p_1 = 1.60$ GeV/c  
do not exceed,
respectively, $q_{max}=217$ MeV/c and $p^{max}_3 = 180$ MeV/c which is
in  complete accordance with the underlying assumptions.

 The FSI corrections calculated for $J_{max}=$1 and 2,
are denoted by the dash- dotted and solid lines, respectively. The agreement between
them is quite satisfactory, especially for the differential cross section
(Fig. 7), where the lines for   $J_{max}=$1 and 2 are indistinguishable.
As before, the PWIA predictions are denoted by dashed lines in all the 
figures (Figs.7-11).

Unlike the picture in Fig. 2,
the $p_1$-dependence of the reaction cross section displayed for
 kinematics II in Fig. 7 covers the QFP region up to its boundary,
i.e., the maximum $p_1$  value
is allowed under the condition $p_r = 0$.
There has been the expectation that the FSI influence is negligible in the QFS
region. However, as  is evident from Fig. 7, the differential cross section is 
$\sim 2$ times smaller relative to the PWIA predictions in the QFS region due to
 the FSI effects. It is the consequence of the vicinity  of 
the two
peaks (QFS and FSI). It is interesting to note that the
PWIA and FSI predictions  for the polarization observables
(Figs.8-11) in the QFS region  practically coincide while these observables change
 their behaviors very quickly when going to the FSI region.

\section{Conclusion}

In the present paper we have explored a reasonable approximation to
treat 
the deuteron-proton breakup reaction in the GeV-region. The employed
approach incorporates
the leading terms in the multiple scattering expansion of
the reaction transition operator in the $N-N$ t-matrix. Our consideration is
free from  shortcomings made in earlier theoretical studies of
this process.
The differential
cross sections and the first-order polarization observables have been
calculated under the two kinematic conditions (one of them corresponding to
 the Gatchina experiment).
These calculations have shown that the FSI contributions are very important
to  describe
 deuteron breakup by the proton in the GeV-region. We have seen that 
the FSI effects,  dependent on kinematic conditions,
appear in different ways. They can slightly change the corresponding
cross section with unpolarized observables and vice versa.
Therefore,  studying the deuteron structure (especially, the
non-nucleonic degrees of freedom in the deuteron) by high- energy breakup reactions, 
one should make every effort to include  FSI
corrections.

\begin{acknowledge}
The authors are thankful to V.P. Ladygin for helpful discussions and permanent
 interest in this work. They thank also O.G.Grebenyuk for information about
 experimental data.
  A.S. is very grateful to V.V. Glagolev for  support and
 hospitality during his visits to the LHE of the JINR,Dubna.

\end{acknowledge}

\appendix
\section*{Appendix}

The NN amplitudes  $F_i$ in the laboratory frame are related to A-F, the ones in
the center-of-mass system by
\begin{eqnarray}
F_1&=&A+B{U^2\over {K_1K_2}}+iCU\left ({1\over {K_1}}-{1\over {K_2}}
\right )
\nonumber\\
F_2&=&iU\left ({A\over {K_1}}-{B\over {K_2}}\right )+
C\left (1+{U^2\over {K_1K_2}}\right )
\nonumber\\
F_3&=&-iU\left ({A\over {K_2}}-{B\over {K_1}}\right )+
C\left (1+{U^2\over {K_1K_2}}\right )
\nonumber\\
F_4&=&D \left\{ (\hat q^*\vec x)-{U\over {K_1}}(\hat q^*\vec z)\right\}
\left\{ (\hat q^*\vec x)+{U\over {K_2}}(\hat q^*\vec z)\right\}+
F\left\{ (\hat Q^*\vec x)-{U\over {K_1}}(\hat Q^*\vec z)\right\}
\left\{ (\hat Q^*\vec x)+{U\over {K_2}}(\hat Q^*\vec z)\right\}
\nonumber\\
F_5&=&A{U^2\over {K_1K_2}}+B+iCU\left ({1\over {K_1}}-{1\over {K_2}}\right )
\nonumber\\
F_6&=&D \left\{ (\hat q^*\vec z)+{U\over {K_1}}(\hat q^*\vec x)\right\}
\left\{ (\hat q^*\vec z)-{U\over {K_2}}(\hat q^*\vec x)\right\}+
F\left\{ (\hat Q^*\vec z)+{U\over {K_1}}(\hat Q^*\vec x)\right\}
\left\{ (\hat Q^*\vec z)-{U\over {K_2}}(\hat Q^*\vec x)\right\}
\nonumber\\
F_7&=&D \left\{ (\hat q^*\vec x)-{U\over {K_1}}(\hat q^*\vec z)\right\}
\left\{ (\hat q^*\vec z)-{U\over {K_2}}(\hat q^*\vec x)\right\}+
F\left\{ (\hat Q^*\vec x)-{U\over {K_1}}(\hat Q^*\vec z)\right\}
\left\{ (\hat Q^*\vec z)-{U\over {K_2}}(\hat Q^*\vec x)\right\}
\nonumber\\
F_8&=&D \left\{ (\hat q^*\vec z)+{U\over {K_1}}(\hat q^*\vec x)\right\}
\left\{ (\hat q^*\vec x)+{U\over {K_2}}(\hat q^*\vec z)\right\}+
F\left\{ (\hat Q^*\vec z)+{U\over {K_1}}(\hat Q^*\vec x)\right\}
\left\{ (\hat Q^*\vec x)-{U\over {K_2}}(\hat Q^*\vec z)\right\}.
\nonumber
\end{eqnarray}
Here we have introduced
\begin{eqnarray}
U&=&\frac {|\vec p\times \vec p_1 |}{\sqrt {s^\prime}}=
\frac {pp_1 sin \theta_1}{\sqrt {s^\prime}}
\nonumber\\
s&=&(p+p^\prime)^2,~~~s^\prime=(p_1+p_2^\prime)^2
\nonumber\\
u_0&=&\frac {E_{\vec p}+m}{\sqrt{s}}~~~
u^\prime_0=\frac {E_{\vec p_1}+E_{\vec p_2^\prime}}{\sqrt{s^\prime}}
\nonumber
\end{eqnarray}
\begin{eqnarray}
K_1={1\over 2}(u^\prime_0+1)(2m+\sqrt{s^\prime})+{1\over 2}(E_{\vec p_1}-E_{\vec p_2^\prime})
\nonumber\\
K_2={1\over 2}(u^\prime_0+1)(2m+\sqrt{s^\prime})-{1\over 2}(E_{\vec p_1}-E_{\vec p_2^\prime}).
\nonumber
\end{eqnarray}

The vectors $\vec q^*$ and  $\vec Q^*$ defined in the c.m.s. can be written in the lab.
system as
\begin{eqnarray}
(\vec q^* \vec x)&=&-(\vec Q^* \vec x)=-p_1 sin \theta_1
\nonumber\\
(\vec q^* \vec z)&=&-p_1 cos\theta_1+p\left\{
\frac {m+\sqrt {s}/2}{\sqrt {s}(1+u_0)}+
\frac {E_{\vec p_1}+\sqrt {s^\prime}/2}{\sqrt {s^\prime}(1+u^\prime_0)}\right\}
\nonumber\\
(\vec Q^* \vec z)&=&p_1 cos\theta_1+p\left\{
\frac {m+\sqrt {s}/2}{\sqrt {s}(1+u_0)}-
\frac {E_{\vec p_1}+\sqrt {s^\prime}/2}{\sqrt {s^\prime}(1+u^\prime_0)}\right\},
\nonumber
\end{eqnarray}
and their moduli are expressed through the invariants  $s$ and $t$ via
\begin{eqnarray}
\vec q^{*2}={1\over 4}(\sqrt {s}-\sqrt {s^\prime})^2-t,~~~
\vec Q^{*2}={1\over 4}(\sqrt {s}+\sqrt {s^\prime})^2+t-4m^2.
\nonumber
\end{eqnarray}

\newpage
{\bf\Large Figures.}\\

{\bf Fig.1} Kinematic variables for the $pd\to ppn$ reaction in the deuteron rest frame.\\

{\bf Fig.2} The breakup differential cross section at $\theta _1 =15^0$,
$\theta _2 = 78^0$ \cite {Bel} plotted versus the  fast proton momentum
$p_1$ in the deuteron rest frame. The dashed line corresponds to the PWIA;
the dotted, dash-dotted and solid curves represent the full calculation with
$J_{max}=$2, 3, and 4, respectively.\\

{\bf Fig.3} The $p_1$-dependence of the vector analyzing power of the initial proton
$C_{0,y,0,0,0}$ at $\theta _1 =15^0$, $\theta _2 =78^0$. Notations as in
Fig.2.\\

{\bf Fig.4} The $p_1$-dependence of the deuteron vector analyzing power
$C_{y,0,0,0,0}$ at $\theta _1 =15^0$, $\theta _2 =78^0$. Notations as in
Fig.2.\\

{\bf Fig.5} The $p_1$-dependence of the deuteron tensor analyzing power $C_{yy,0,0,0,0}$
at $\theta _1 =15^0$, $\theta _2 =78^0$. Notations as in Fig.2.\\

{\bf Fig.6} The $p_1$-dependence of the deuteron tensor analyzing power $C_{zz,0,0,0,0}$
at $\theta _1 =15^0$, $\theta _2 =78^0$. Notations as in Fig.2.\\

{\bf Fig.7} The breakup differential cross section at $\theta _1 =6.77^0$,
$\theta _2 = 83^0$  plotted versus the  fast proton momentum
$p_1$ in the deuteron rest frame. The dashed line corresponds to the PWIA;
the dash-dotted and solid curves represent the full calculation with
$J_{max}=$1 and 2, respectively.\\

{\bf Fig.8} The same as in Fig. 3 at $\theta _1 =6.77^0$, $\theta _2 =83^0$.
Notations  as in Fig. 7.\\

{\bf Fig.9} The same as in Fig. 4 at $\theta _1 =6.77^0$, $\theta _2 =83^0$.
Notations  as in Fig. 7.\\

{\bf Fig.10} The same as in Fig. 5 at $\theta _1 =6.77^0$, $\theta _2 =83^0$.
Notations  as in Fig. 7.\\

{\bf Fig.11} The same as in Fig. 6 at $\theta _1 =6.77^0$, $\theta _2 =83^0$.
Notations  as in Fig. 7.\\

\newpage

\begin{center}
\hspace*{-2.5cm}
\vspace*{-2.5cm}
\unitlength=1.10mm
\special{em:linewidth 0.4pt}
\linethickness{0.4pt}
\begin{picture}(81.33,92.67)
\put(12,41.0){\vector(1,0){6.}}
\put(28.33,41.0){\circle*{2.5}}
\put(28.33,41.){\line(-1,0){30.}}
\put(28.33,41.){\line(1,0){4}}
\put(36.33,41.){\line(1,0){4}}
\put(44.33,41.){\line(1,0){4}}
\put(52.33,41.){\line(1,0){4}}
\put(60.33,41.){\line(1,0){4}}
\put(68.33,41.){\line(1,0){4}}
\put(28.33,41.){\line(2,1){30.00}}
\put(39.33,46.60){\vector(2,1){8.00}}
\put(28.33,41.){\line(4,1){40.00}}
\put(50.,46.60){\vector(4,1){6.00}}
\put(28.33,41.){\line(3,-4){20.00}}
\put(38.,28.20){\vector(3,-4){4.00}}
\put(12,44){\makebox(0,0)[cc]{$\vec{p}$}}
\put(55,50.){\makebox(0,0)[cc]{$\vec{p_1}$}}
\put(40,20.){\makebox(0,0)[cc]{$\vec{p_2}$}}
\put(50,55.){\makebox(0,0)[cc]{$\vec{p_3}$}}
%
\put(40,42.2){\makebox(0,0)[cc]{ $ )$}}
\put(37.5,36.2){\makebox(0,0)[cc]{$\biggr )$}}
\put(45.,43.){\makebox(0,0)[cc]{\large $\theta _1$}}
\put(42.,34.){\makebox(0,0)[cc]{\large $\theta _2$}}


\end{picture}
\end{center}

{\bf a}

\hspace*{-4cm}
\begin{center}
\hspace*{-2.5cm}
\unitlength=1.10mm
\special{em:linewidth 0.4pt}
\linethickness{0.4pt}

\begin{picture}(81.33,92.67)
\put(5.67,40.00){\line(1,0){22.67}}
\put(28.33,40.00){\line(0,1){1.33}}
\put(28.33,41.33){\line(-1,0){22.67}}
\put(61.67,40.67){\line(-1,0){33.00}}
\put(28.67,40.67){\line(0,0){0.00}}
\put(28.67,40.67){\line(3,5){16.00}}
\put(44.67,67.33){\line(3,-5){16.33}}
\put(61.00,40.33){\line(2,1){20.33}}
\put(61.00,40.67){\line(5,-2){20.33}}
\put(45.00,67.00){\line(2,1){21.33}}
\put(44.33,67.33){\line(-2,1){21.67}}
\put(44.33,67.33){\line(-2,1){21.33}}
\put(45.00,67.00){\line(-2,1){21.33}}
\put(28.33,41.00){\circle*{3.33}}
\put(61.00,40.67){\circle*{3.33}}
\put(45.00,67.00){\circle*{3.33}}
\put(69.33,44.60){\vector(2,1){4.00}}
\put(70.00,37.33){\vector(2,-1){3.00}}
\put(52.67,70.67){\vector(2,1){6.00}}
\put(29.33,75.00){\vector(2,-1){6.00}}
\put(58.00,68.00){\makebox(0,0)[cc]{$\vec{p_1}$}}
\put(31.67,68.67){\makebox(0,0)[cc]{$\vec{p}$}}
\put(77.00,51.67){\makebox(0,0)[cc]{$\vec{p_2}$}}
\put(76.67,28.67){\makebox(0,0)[cc]{$\vec{p_3}$}}
\put(30.33,53.33){\makebox(0,0)[lc]{$\vec{p^\prime}$}}
\put(57.67,52.33){\makebox(0,0)[lc]{$\vec{p_2^\prime}$}}
\put(40.33,34.67){\makebox(0,0)[lc]{$\vec{p_3^\prime}$}}
\end{picture}
\end{center}

\vspace{-3cm}
{\bf b}

\vspace *{3cm}
{\bf Fig.1}
\newpage
\vspace*{20cm}

\begin{figure}[h]
\includegraphics{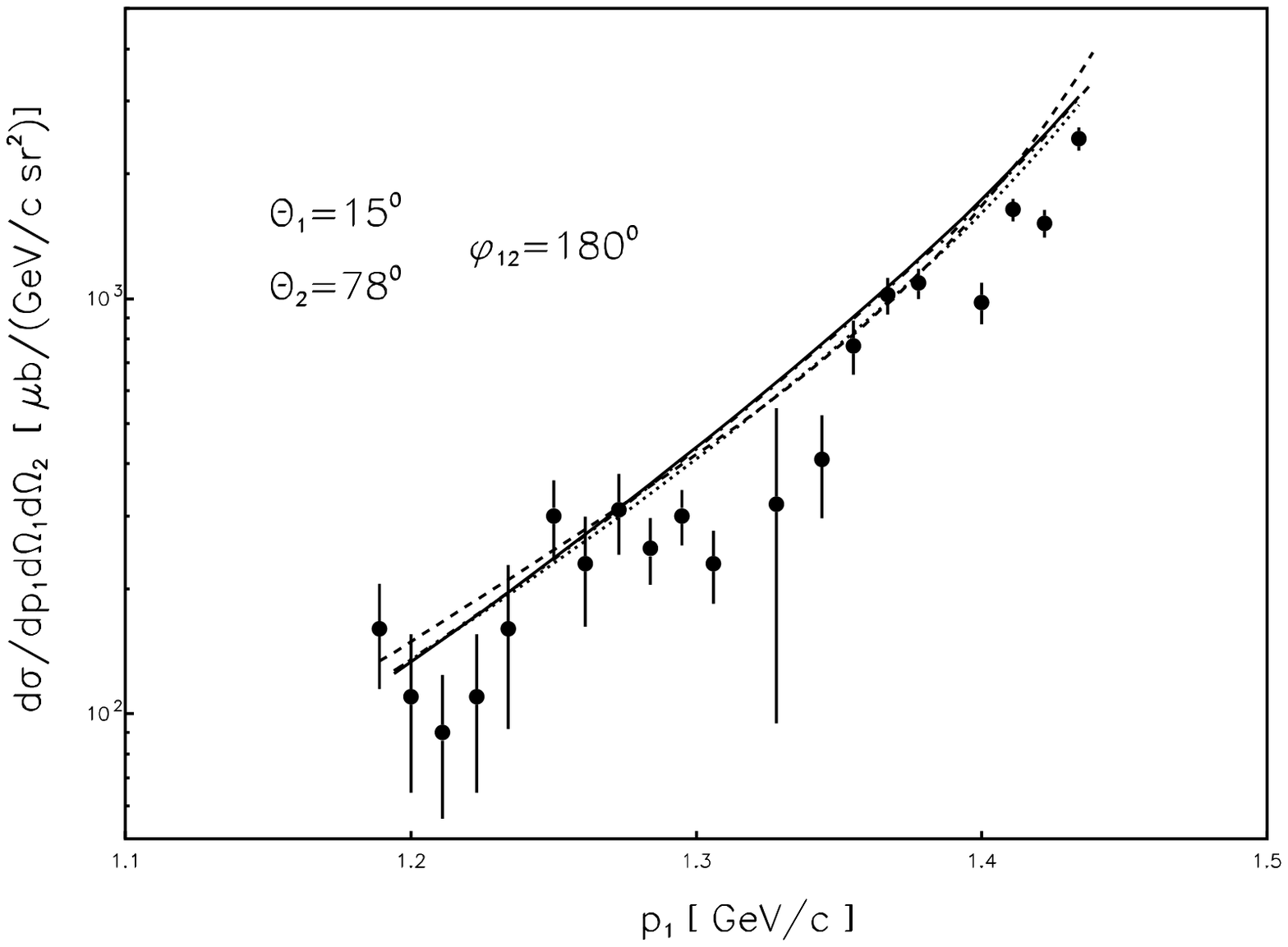}
\end{figure}
\vspace*{-5cm}
{\bf Fig.2}
\newpage
\vspace*{20cm}

\begin{figure}[h]
\includegraphics{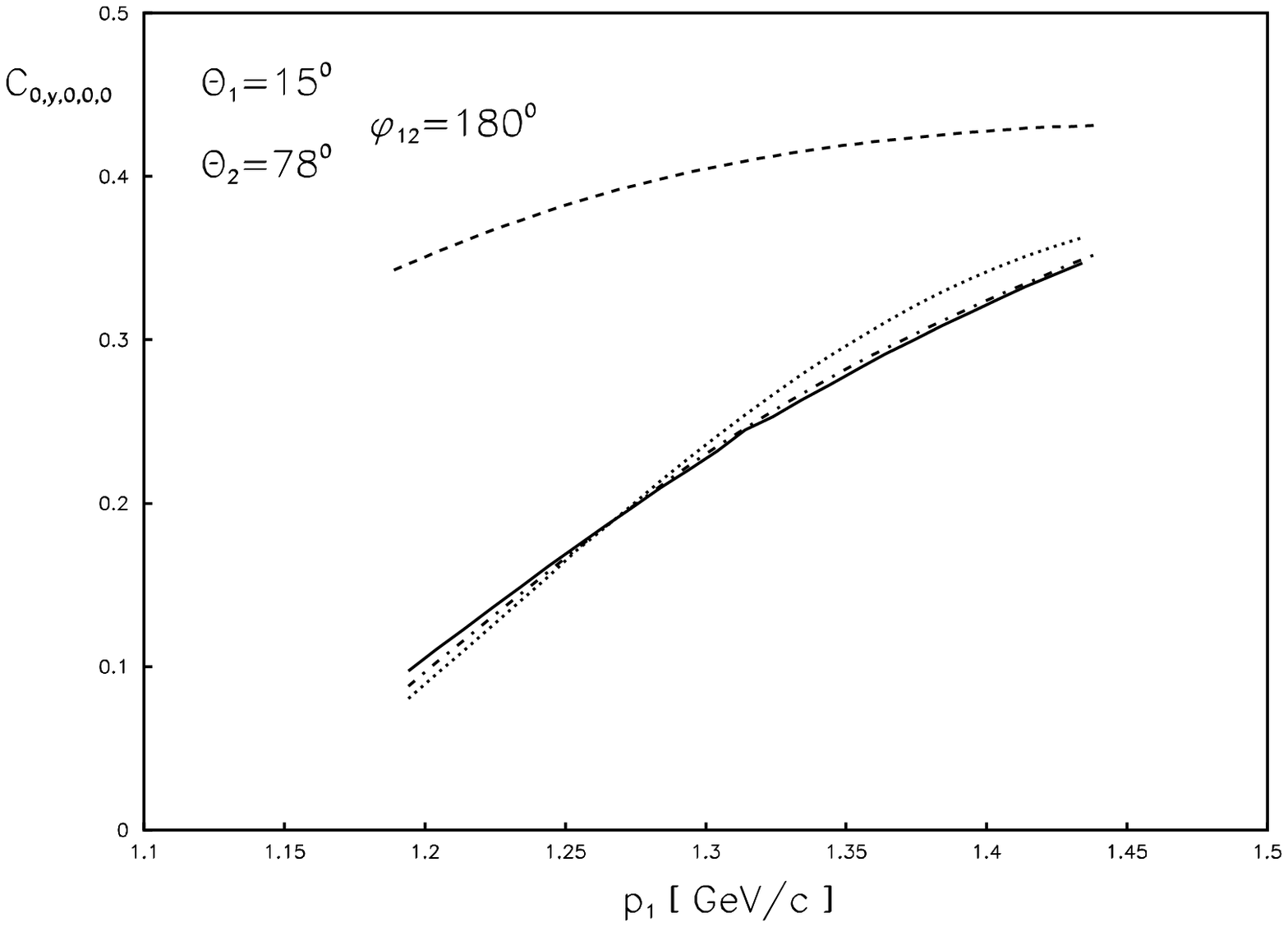}
\end{figure}
\vspace*{-5cm}
{\bf Fig.3}
\newpage
\vspace*{20cm}

\begin{figure}[h]
\includegraphics{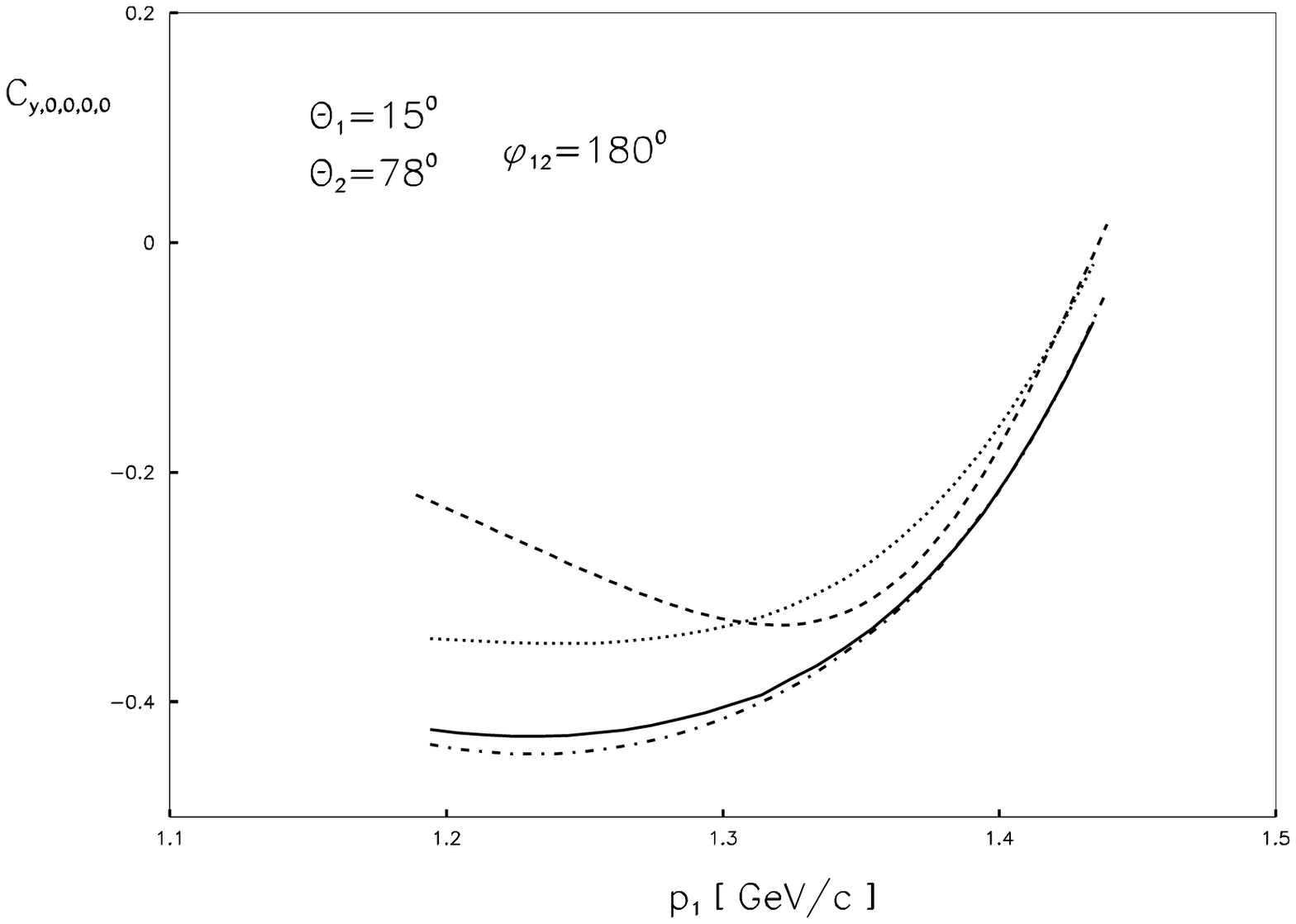}
\end{figure}
\vspace*{-5cm}
{\bf Fig.4}
\newpage
\vspace*{20cm}

\begin{figure}[h]
\includegraphics{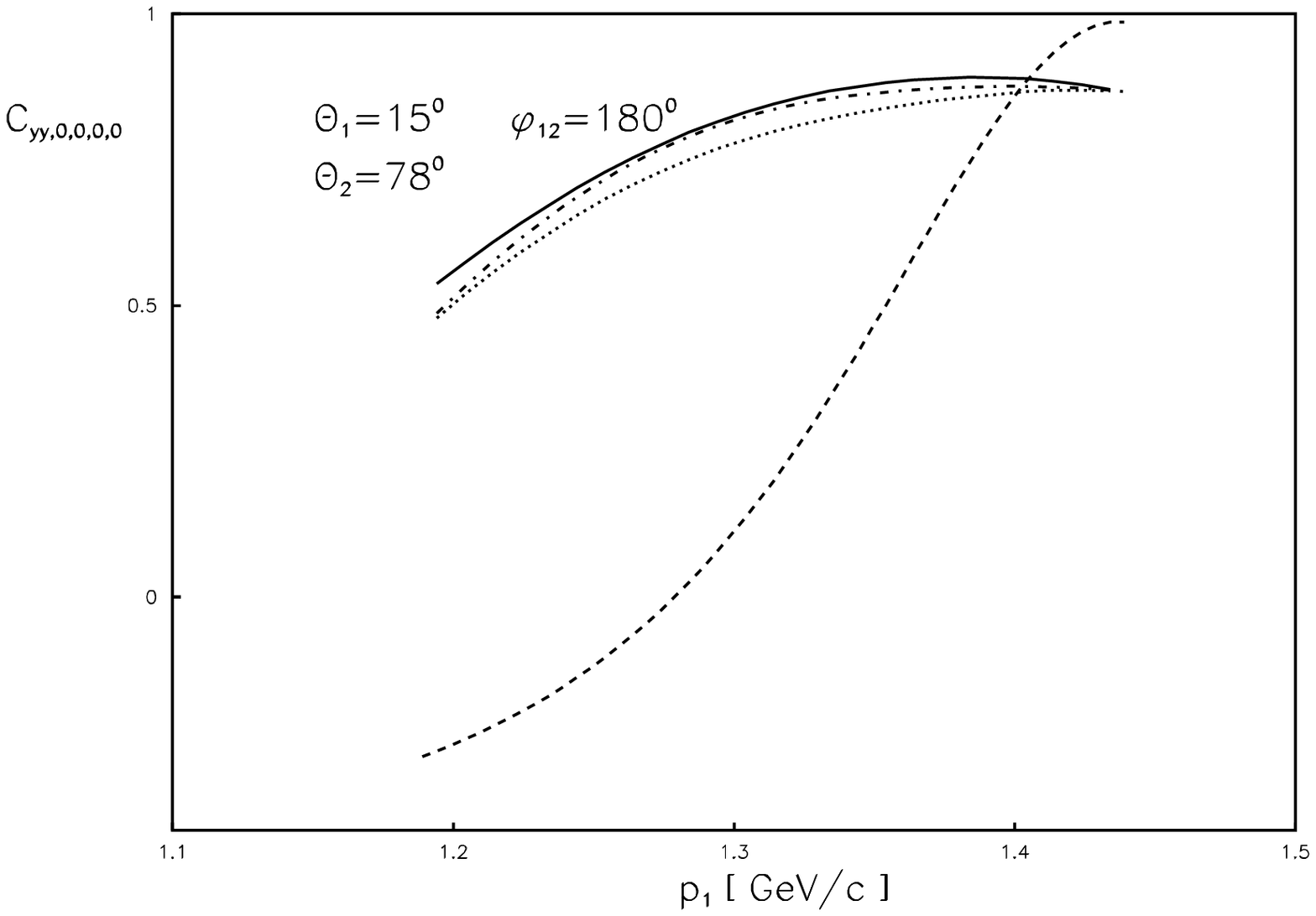}
\end{figure}
\vspace*{-5cm}
{\bf Fig.5}
\newpage
\vspace*{20cm}

\begin{figure}[h]
\includegraphics{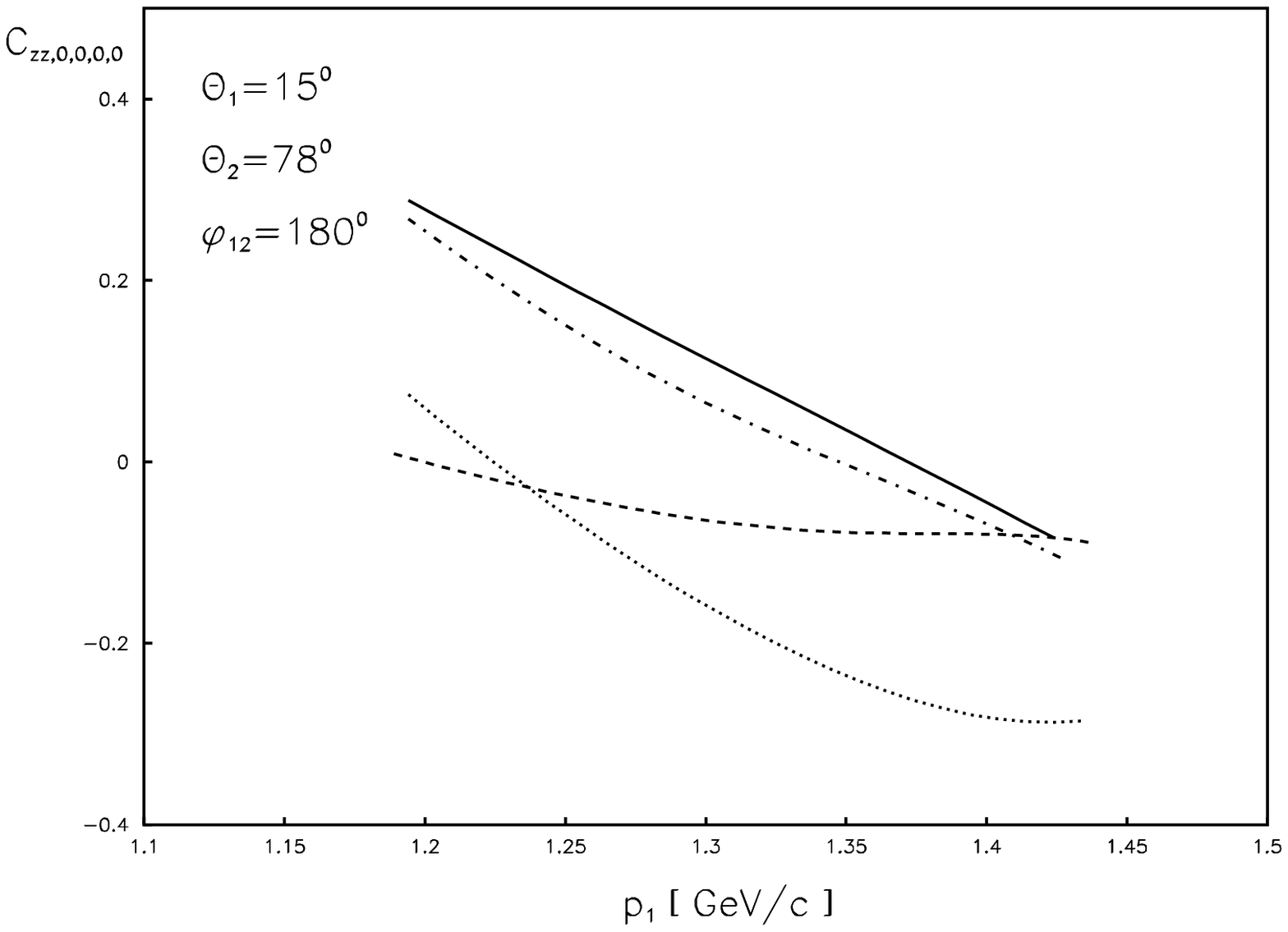}
\end{figure}
\vspace*{-5cm}
{\bf Fig.6}
\newpage
\vspace*{20cm}

\begin{figure}[h]
\includegraphics{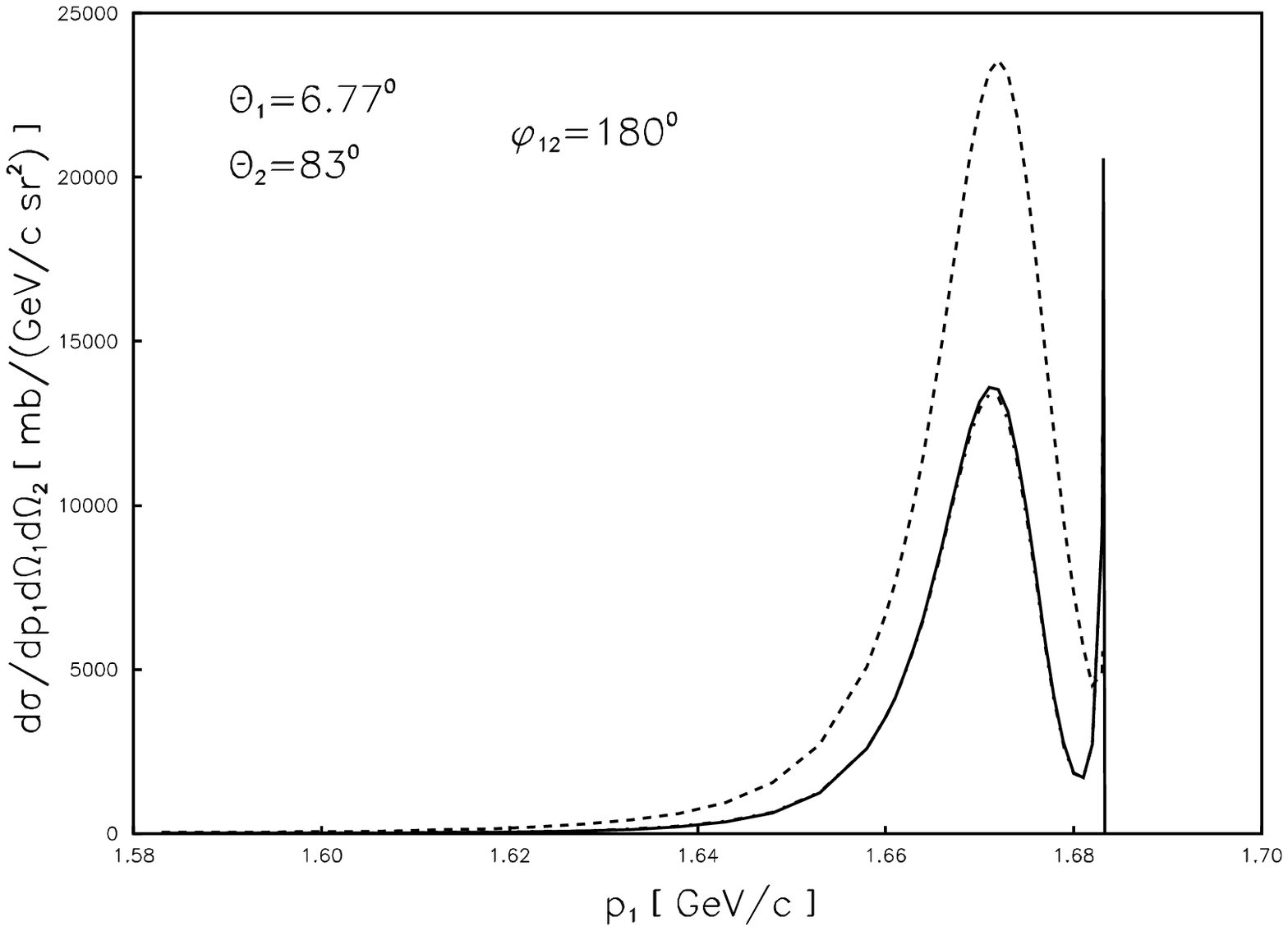}
\end{figure}
\vspace*{-5cm}
{\bf Fig.7}
\newpage
\vspace*{20cm}

\begin{figure}[h]
\includegraphics{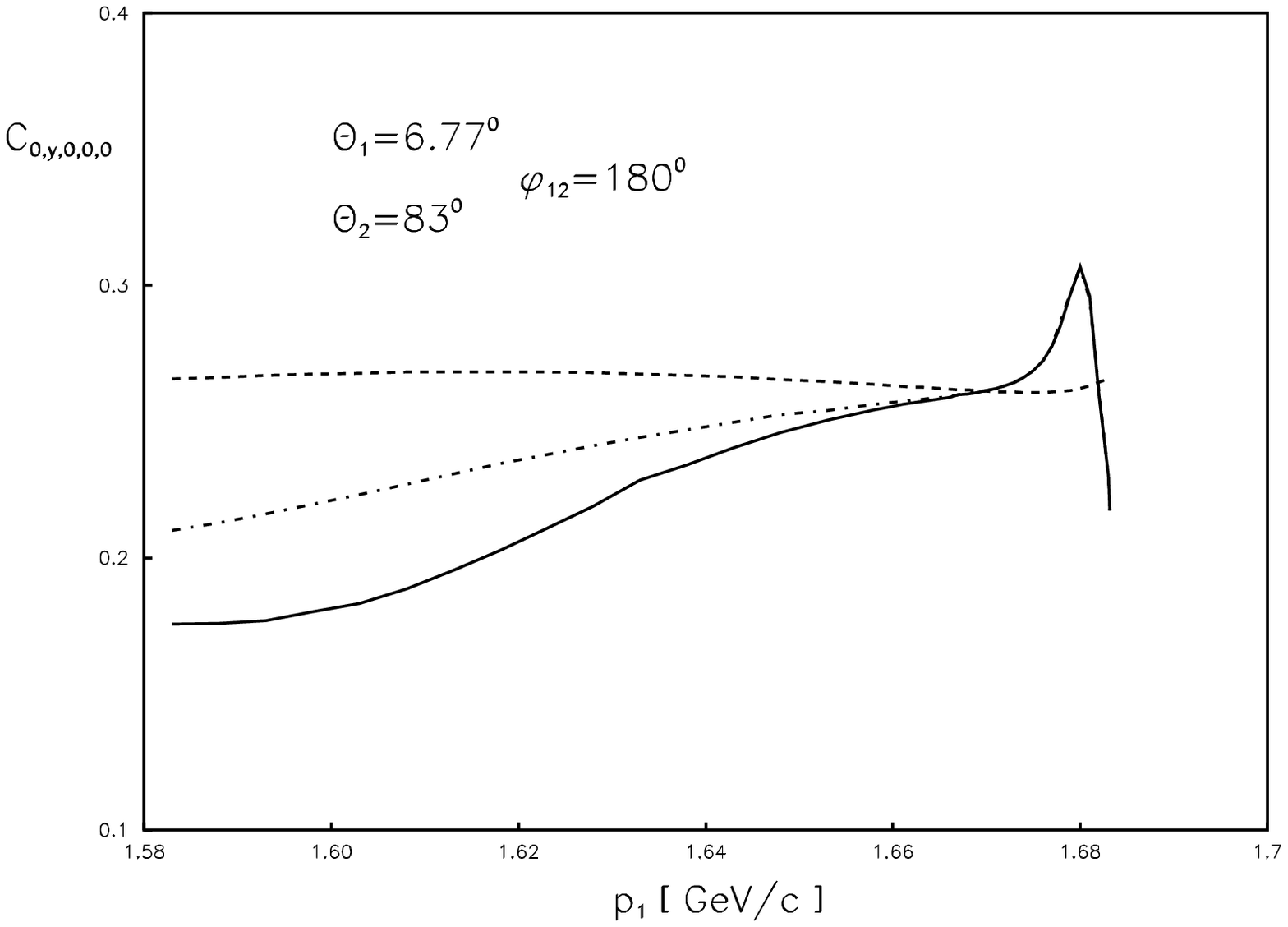}
\end{figure}
\vspace*{-5cm}
{\bf Fig.8}
\newpage
\vspace*{20cm}

\begin{figure}[h]
\includegraphics{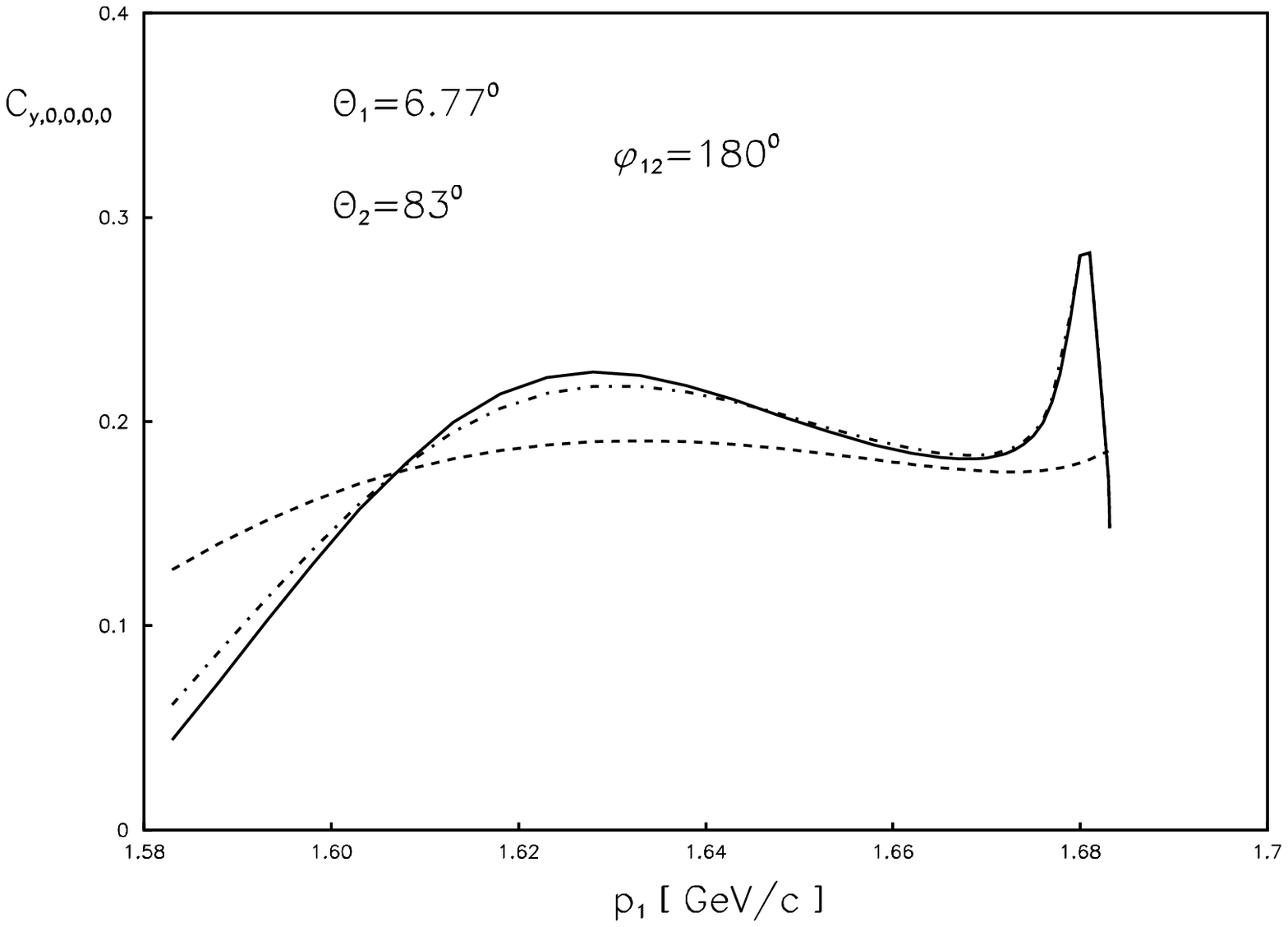}
\end{figure}
\vspace*{-5cm}
{\bf Fig.9}
\newpage
\vspace*{20cm}

\begin{figure}[h]
\includegraphics{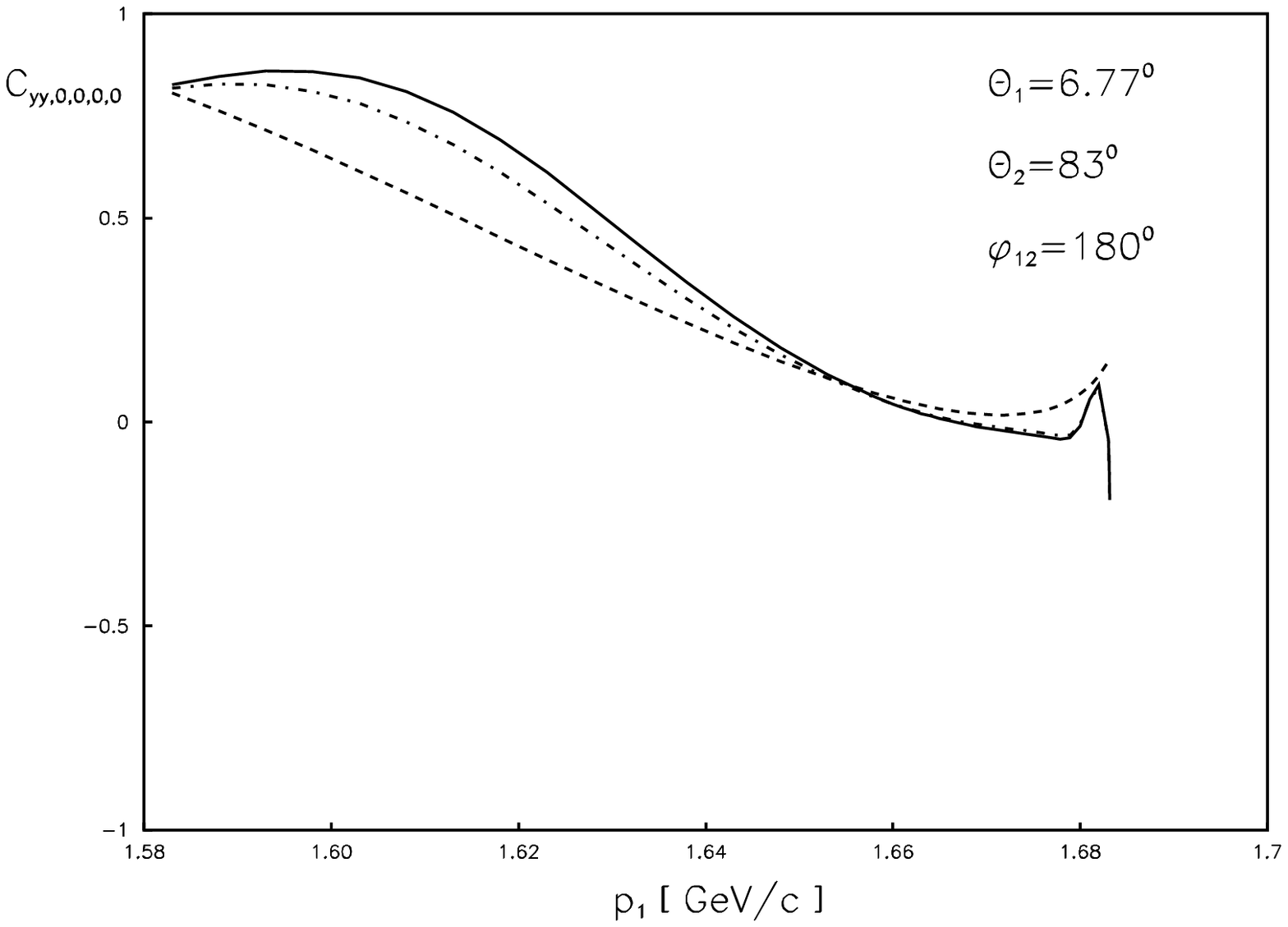}
\end{figure}
\vspace*{-5cm}
{\bf Fig.10}

\newpage
\vspace*{20cm}

\begin{figure}[h]
\includegraphics{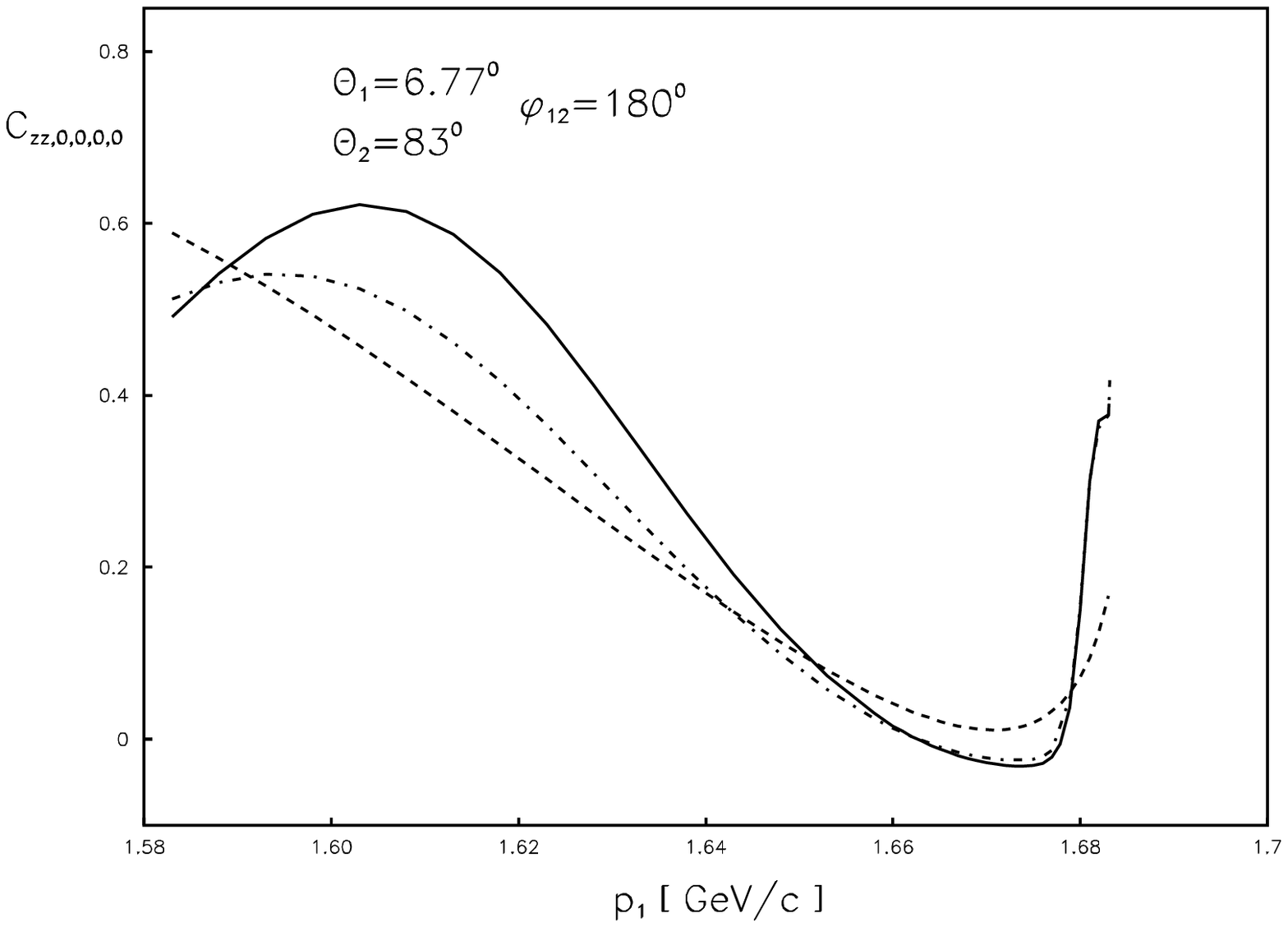}
\end{figure}
\vspace*{-5cm}
{\bf Fig.11}

\end{document}